\documentstyle[seceq,mbf,epsf,wrapft]{ptptex}

\pubinfo{Vol.~ **, No.~**, January  2002}  
\preprintnumber[3cm]{
KUNS-1325\\PTPTeX ver.0.8\\ August, 1997}

\title{
Dynamical Effects beyond Mean Field  in Drip Line Nuclei
}

\author{
Hiroyuki  {\sc Sagawa}\footnote{ E-mail address:sagawa@u-aizu.ac.jp}
 Gianluca   {\sc Col\`o}$^{1}$
  and Toshio  {\sc Suzuki}$^{2}$
}

\inst{
   Center for Mathematical Sciences, 
                  University of Aizu,  \\
                  Aizu-Wakamatsu, Fukushima 965-8580,
                  Japan\\

$^{1}$ Dipartimento di Fisica, Universit\`a degli Studi 
and INFN \\
Sezione di Milano, Via Celoria 16, 20133 Milano, 
Italy\\

$^{2}$Department of Physics, College of
Humanities and Sciences, Nihon University\\
Sakurajosui 3-25-40, Setagaya-ku, Tokyo 156, 
Japan
}

\recdate{
January 11, 2002}

\abst{
New  structure problems are
raised in nuclei near drip lines by recent RIB experiments.
We pointed out  the importance of dynamical effect 
beyond mean field theories to  the new  structure problems.
}

\begin{document}

\maketitle


\makeatletter
\if 0\@prtstyle
\def\asp{.3em} \def\bsp{.26em}
\else
\def\asp{.3em} \def\bsp{.3em}
\fi \makeatother

\section{Introduction}
Recently, nuclear structure in neutron-rich nuclei has attracted 
much attention because of its exotic nature, such as halo and skin
\cite{Tan92,HJJ95,HJ87}.  
Recent advances of secondary beam techniques allow us to measure useful 
structure information 
on masses, interaction cross sections and 
excited states. The shell structure is one of essential issues in nuclear 
structure. Recently, the magic numbers in the neutron-rich region have been 
extensively studied both experimentally and theoretically.
For example, the melting of the N=8 closed shell 
was pointed out due to the large mixing of 1p$_{1/2}$ and 2s$_{1/2}$ 
orbitals in $^{11}$Be \cite{Simon99},  $^{11}$Li \cite{Keller94} and
$^{12}$Be \cite{Iwa00}. The disappearance of the N=20
closed shell was also 
shown in $^{32}$Mg experimentally \cite{Mg32}. 
Very recently a possible new magic number 
is pointed out at N=16 by the analysis of the neutron separation energies 
and the interaction cross sections \cite{Ozawa00}. 

So far the nuclear structure  problems in light nuclei
   have been studied mostly 
by using mean field theory \cite{HF} and shell model calculations 
\cite{shell}. 
It is known that the correlations beyond the mean field
have substantial effects on the 
single-particle energies and electromagnetic transitions. 
 In particular, the effect of collective
excitations 
has been studied by 
the particle-vibration coupling and found to be important\cite{BM75,Mah85}. 
In this paper, we present the study of dynamical correlations on
   single-particle energies in light 
neutron-rich nuclei based on the $^{10}$Be and $^{24}$O core by using 
the particle-vibration coupling model\cite{Colo01}. 
The polarization charges for electric transitions in drip line nuclei 
  have been discussed also by using the particle-vibration coupling model
 elsewhere\cite{HSZ97,SA01}.
   In the 
present version of the model, firstly  
Hartree-Fock (HF) calculations with Skyrme interactions 
are performed to obtain the zeroth-order single-particle 
energies and the wave functions. Secondly, we couple to them 
the natural parity vibrational states
with J$^{\pi}$=1$^-$, 2$^+$ and 3$^-$ by using the
particle-vibration coupling Hamiltonian derived from the same 
Skyrme interaction, and the HF wave functions.  
In Section 2, we 
describe our model. Numerical results are given in Section 3.
Summary is given in Section 4.

\section{Particle-vibration coupling model}

\begin{wrapfigure}{r}{6.6cm}
            \parbox{\halftext}{
 \epsfxsize = 6.6 cm 
\epsfbox{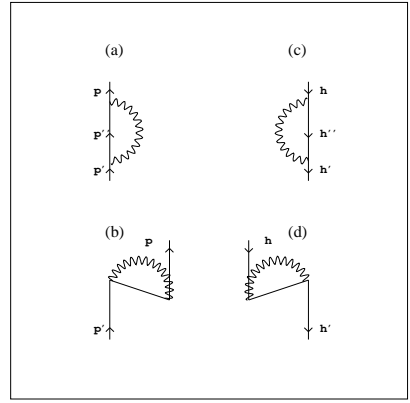}
        \caption{The four diagrams corresponding to 
 the second-order couplings of 
particles
(left column) or holes (right column) with vibrations, represented by wavy
lines. See the text    for details. }
        \label{fig:1}}

\end{wrapfigure}

The particle-vibration coupling Hamiltonian has 
the form
\begin{equation}
 H_{PV} = \sum_{\alpha\beta} \ \sum_{\lambda n \mu} \langle \alpha |
 \delta\varrho_n (r) v(r) Y_{\lambda\mu}(\hat r) | \beta \rangle
 \ a^\dagger_\alpha a_\beta,
\label{pvc}\end{equation}
where $\alpha,\beta$ label the single-particle states, $v(r)$ is 
the form factor associated to 
the particle-hole interaction 
$V_{ph}$
derived from the Skyrme force 
($V_{ph}(\vec r_1,\vec r_2)=v(r_1)\delta(\vec r_1 -\vec r_2)$~\cite{note}), 
and the 
radial transition density $\delta\varrho_n(r)$ 
can be derived from
the RPA 
phonon.  

The  second-order correction due to the RPA phonons to the single-particle
energies coming from the perturbing Hamiltonian (\ref{pvc}) 
is obtained by
the real part 
of the four diagrams depicted in Fig. 1. 
The imaginary part, which gives rise to the width of the single-particle
states, is not discussed in the present work.
The corresponding energy shifts for the diagrams can 
be found, e.g., in Ref.~\cite{VG80}, and are given to be, 
\begin{eqnarray}
\Delta\varepsilon_\alpha = Re\ \sum_{\lambda,n} {1\over 2j_\alpha+1}
[
& \sum_{\beta > F} & {\vert \int dr u_\alpha(r) u_\beta(r) v(r)
\delta\varrho_n(r) \vert^2 \vert \langle \alpha \vert\vert Y_\lambda
\vert\vert \beta \rangle \vert^2 \over \varepsilon_\alpha-(\varepsilon_\beta
+ \omega_n) + i\eta} + \nonumber \\
& \sum_{\beta < F} & {\vert \int dr u_\alpha(r) u_\beta(r) v(r)
\delta\varrho_n(r) \vert^2 \vert \langle \alpha \vert\vert Y_\lambda
\vert\vert \beta \rangle \vert^2 \over \varepsilon_\alpha-(\varepsilon_\beta
- \omega_n) - i\eta} ],
\label{eq:2} 
\end{eqnarray} 
where $\varepsilon_\alpha$ and $\varepsilon_\beta$ are HF single-particle
energies, $u_\alpha$ and $u_\beta$ are the radial HF wave functions
and $\omega_n$ is the RPA phonon energy.
The sum over $\beta$ spans unoccupied particle states (occupied hole 
states) when the notation $\beta > F$ ($\beta < F$) is employed. 
If the state $\alpha$ is a particle state, the first and second terms in the
above equation correspond to the upper and lower left-hand side diagrams
 (a) and (b) in 
Fig. 1. If $\alpha$ is 
a hole state, the first and second terms correspond,
respectively, to the lower and upper right-hand side diagrams (d) and (c) 
in Fig. 1. On quite general grounds, it is expected that the real parts 
of Eq. (\ref{eq:2})  are
larger for the ``leading'' diagrams (a) and (c) shown in Fig. 1 
than the  diagrams (b) and (d) in Fig. 1. This is because the
energy denominators of the upper row 
  diagrams (a) and (c)   
are expected to be smaller than those of the  diagrams (b) and (d). 
Furthermore, the energy denominator of the contributions (b) and (d)
have opposite sign, so
their contributions  should cancel partially those  of the leading terms. 
In $^{24}$O we will find exceptions to these
rules 
due to the possibility that the leading contributions
are not possible because of phase space limitations (hole intermediate states
may not be available in these light nuclei if large angular momenta are
required by selection rules). This will be one of the important qualitative 
messages of the present work.

\begin{wrapfigure}{r}{6.6cm}
            \parbox{\halftext}{
           \epsfxsize = 6.6 cm   
\epsfbox{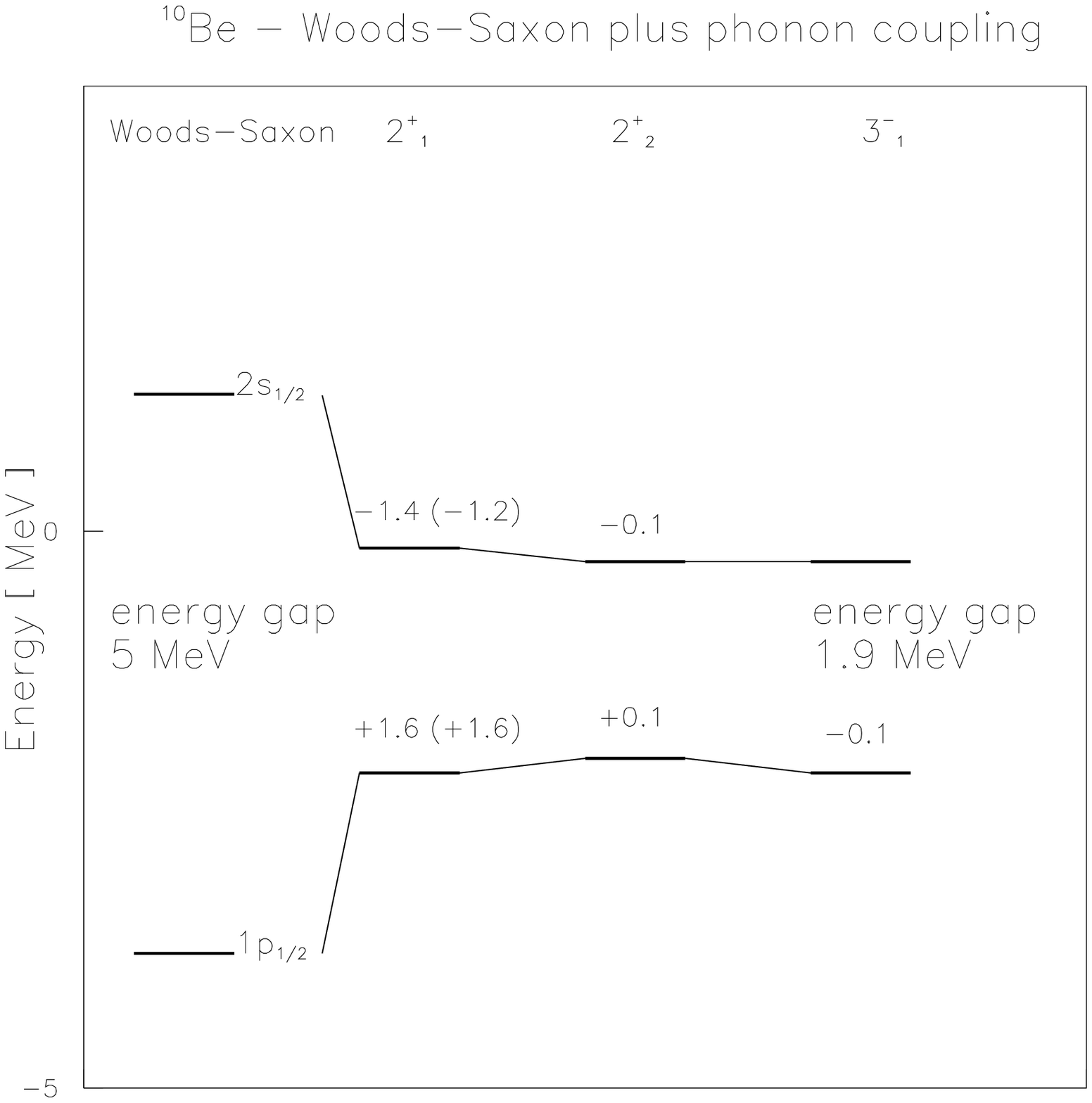}
        \caption{The neutron single-particle spectrum
 in $^{10}$Be core calculated by
  a Woods-Saxon potential plus particle-vibration coupling. The energies are
obtained, following the levels from left to right, by  the Woods-Saxon 
potential, and including the effect of the 2$^+_1$ state,
 of the  2$^+_1$ and 2$^+_2$ states,
and of the 2$^+_1$,  2$^+_2$  and 3$^-_1$ states, respectively.
  The excitation energies and $B(E2)$ values of the
2$^+_1$ and 2$^+_2$ states are taken from experimental data~\cite{AS}, while
those of the 3$^-_1$ state are obtained by shell model calculations. The
cutoff radius for the calculation of the single-particle wave functions is
taken to be 20 fm. Values corresponding to 15 fm are given in parenthesis for
the sake of comparison. See the text for more details.}
        \label{fig:2}}
       \end{wrapfigure}

\section{Results}

We present numerical results for the single-particle energies in 
the $^{10}$Be
and $^{24}$O core. 
The shell structure of $^{10}$Be core in the present study is 
compared   with the 
results of Ref.~\cite{NVM} (see also the coupled-channel calculation of 
Ref.~\cite{Nun96}). 
Since it is somewhat questionable to apply the  self-consistent HF+RPA
approach to obtain realistic single-particle energies and collective
excited states in very light nuclei, 
we adopt a phenomenological approach for $^{10}$Be.
Namely, we start from a Woods-Saxon potential as the zeroth-order
potential and calculate the particle-vibrational coupling with 
empirical phonon energies and transition strengths.

In $^{10}$Be, the existing Skyrme forces 
give a HF single-particle spectrum with a large 
1p$_{1/2}$-1p$_{3/2}$ gap.  
The interaction SIII gives, for instance, 
$\varepsilon({\rm 1p}_{1/2})$=-3.80 MeV and 
$\varepsilon({\rm 1p}_{3/2})$=-11.13 MeV. Using a standard Woods-Saxon 
parameterization~\cite{BM}, 
\begin{equation}
U = (-V_0 + V_\tau {N-Z\over A}) f(r) + V_{ls} (\vec l\cdot \vec s) 
R_0^2 {f^\prime(r)\over r}
\end{equation}
(with $V_0$ = 55 MeV, $V_\tau$ = 33 MeV, $V_{ls}$ = $\alpha_{ls}V$ and
$\alpha_{ls}$ = 0.44, and where $f(r)$ is the usual Fermi function with 
$r_0$ = 1.27 fm and $a$ = 0.67 fm),  
the energy gap is reduced but the Fermi surface lies in a similar
position as in the Skyrme-HF case. In fact, the Woods-Saxon results are 
$\varepsilon(1p_{1/2})$=-3.93 MeV and 
$\varepsilon(1p_{3/2})$=-9.36 MeV. This 1p$_{1/2}$ single-particle state, 
together with the 2s$_{1/2}$ unoccupied state, are shown in the first column of 
Fig. 2. 

The vibrational states included in our model space are limited to those 
for which  
experimental information is available. The
 energy and reduced transition probability $B(E2, 
0\rightarrow 2^+_1)$
 of the lowest 2$^+_1$ state 
in $^{10}$Be
has been measured  to be 
3.37 MeV and 52 e$^2$fm$^4$, respectively \cite{raman87}.
  The effect of coupling between the 
single-particle states and  this phonon can be seen in the second column of 
Fig. 2. The coupling reduces the 2s$_{1/2}$-1p$_{1/2}$ gap from 5 to 2 MeV, 
melting the N=8 shell closure. Interpreting those as the ${1\over 2}^+$ and 
${1\over 2}^-$ levels of $^{11}$Be, the shifts induced by phonon coupling 
are not enough to give an inversion of the two levels. This is at variance with
Ref.~\cite{NVM},
in which the inversion of the two levels
was obtained by a particle-vibration coupling model
with   a non-standard 
Woods-Saxon parameterization for the mean field: 
a large 
diffuseness
$a$=0.9 fm has been employed and the spin-orbit parameter
$\alpha_{ls}$ has been changed from 0.44 to 0.8 just for p-states.
Thus, in such a modified Woods-Saxon potential, 
the 1p$_{1/2}$ and 2s$_{1/2}$ 
states lie close enough so
that the 
shift induced by phonons can reverse their order. 
      \begin{wrapfigure}{r}{6.6cm}
            \parbox{\halftext}{
       
            \epsfxsize =6.6cm   
\epsfbox{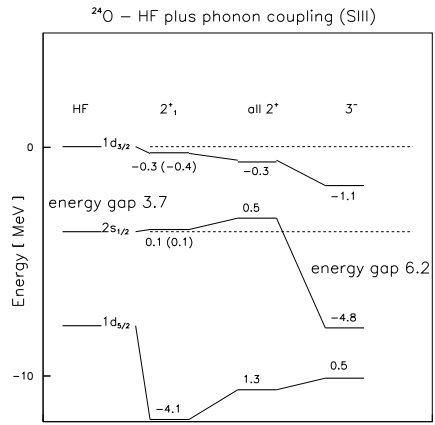}
        \caption{The neutron single-particle spectrum in $^{24}$O core
   calculated with 
SIII-HF plus particle-vibration coupling.  The energies are
obtained, following the levels from left to right, by  the HF  
potential, and including the effect of the 2$^+_1$ state,
 of all  2$^+$ states,
and of all  2$^+$   and 3$^-$ states, respectively.
   The energies and $B(E\lambda)$ values of the excited states 
are calculated by the self-consistent HF+ RPA model. 
See   the text for details. 
 }
        \label{fig:3}}
        \end{wrapfigure}

We checked the above perturbative results for the ${1\over 2}^+$ state by 
 diagonalizing the Hamiltonian~(\ref{pvc}) in the 
five-dimensional model space with  2s$_{1/2}$ ,
1d$_{5/2}\otimes$2$^+$, 2d$_{5/2}\otimes$2$^+$, 1d$_{3/2}\otimes$2$^+$,
2d$_{3/2}\otimes$2$^+$ configurations and found 
  consistent results with the above calculations.
The wave function of the ground state is essentially given by
\begin{equation}
\vert {1\over 2}^+ \rangle = 0.921 \vert \rm 2s_{1/2} \rangle +
                             0.356 \vert \rm 1d_{5/2}\otimes 2^+ \rangle.
\end{equation}

In the case of $^{24}$O core, 
the HF spectrum with SIII interaction is shown in the first 
column of Fig. 3. 
We have then
calculated   0$^+$, 1$^-$, 2$^+$, 3$^-$ and
4$^+$ vibrational states by the  self-consistent RPA model. 
We first check the numerical accuracy of the model.
The energy and proton (neutron) moments $M_p$ ($M_n$) of the
collective lowest 2$^+$ state are 
obtained by three different methods for the RPA calculations.
The energy and proton (neutron) moments are 
 $E$=3.90 MeV and  $M_p$ ($M_n$)=1.8 (10) fm$^2$, respectively,
  with unoccupied levels calculated using the
harmonic oscillator basis

In the case of the coupling with the lowest 2$^+_1$ state, the 1d$_{3/2}$ and
2s$_{1/2}$ states have moderate shifts. 
Although the phonon is 
collective, there is 
substantial cancellation, in the case of the 2s$_{1/2}$ hole, 
between the  diagrams (c) and (d) in  Fig. 1  with the 
1d$_{5/2}$ and 1d$_{3/2}$ intermediate state, respectively.
In the case of the 
1d$_{3/2}$ particle the cancellation occurs between the two 
 diagrams (a) and (b) with the 1d$_{3/2}$ 
and 2s$_{1/2}$ intermediate states, respectively. 
 There is not such cancellation for the 
1d$_{5/2}$ state because the
 diagram (c) of Fig. 1, with intermediate 
2s$_{1/2}$ state, gives a large effect and has negative sign, contrary to the 
general rule.
The remaining quadrupole states have non-negligible effect although they are 
   lying
at higher energies. The 3$^-$ states play a crucial role as far as the
single-particle energy gap is concerned and deserves a special discussion.
The large amount of low-energy (i.e., below 10 MeV) 3$^-$ strength explains 
the large absolute values of the energy shifts. In particular, two states
at 6.07 MeV and at 7.33 MeV absorb 
22\% and 19\% of the energy weighted sum rule 
strength, respectively.  
The fact that the 2s$_{1/2}$ hole state is pushed down in energy is due
to the absence of contributions from the
 diagram (c) of Fig. 1
because of angular momentum selection rules, and 
also due to the strong contribution
from the  diagram (d) with the resonant-like 1f$_{7/2}$ intermediate
state. These unusual  effects are possible only in light nuclei where occupied
states span only low values of the angular momentum, and were never found
in systematic studies of heavier nuclei, for example, near $^{208}$Pb.  It is also
   interesting to
observe  experimentally 
 the predicted collective  3$^-$ states in  $^{24}$O to test the 
validity of  
the present theoretical study of the shell gaps.

\section{Summary}

Effects of the collective modes on the shell structure have been investigated 
in the nuclei 
near the $^{10}$Be and $^{24}$O core, at the neutron drip line. 
Energy shifts of the single-particle states are calculated by the 
particle-vibration coupling model taking into account the coupling to low-lying
collective vibrational states.
In the case of $^{10}$Be core, 
the coupling to the lowest 2$^{+}$ state is found to be the most
important effect to lower and raise the single-particle energies of 2s$_{1/2}$
and 1p$_{1/2}$ states, respectively, and leads to a much narrower gap between
the two states than that of the Woods-Saxon potential. 
In the case of $^{24}$O core, the single-particle energies below the Fermi level 
behaves unusually in the calculations with the particle(hole)-vibration
coupling compared with heavy nuclei near the $^{208}$Pb core. The 
single-particle energy of the 1d$_{5/2}$ state is
lowered by the coupling to the collective 2$^{+}$
  states which gives an upward shift in heavy 
nuclei.  Moreover the energy  of 2s$_{1/2}$ state is also 
lowered by the coupling to 3$^{-}$ states due to 
very specific effects found in the case of $^{24}$O core, i.e., 
  the blocking of the
 available configuration space for the particle(hole)-vibration couplings.

\section*{Acknowledgments}
This  work is supported in part by the Japanese Ministry of Education, 
Science, Sports and Culture by Grant-In-Aid for Scientific Research 
under the program number C(2)  12640284.


\begin{thebibliography}{99}
\bibitem{Tan92}
I. Tanihata et al., Phys. Lett. {\bf B287} (1992) 307, and
references therein;\\
I. Tanihata, J. Phys. {\bf G22} (1996) 157.  
\bibitem{HJJ95}
P. G. Hansen, A. S. Jensen and B. Jonson, Ann. Rev. Nucl. 
Part. Sci. {\bf 45} (1995) 591.
\bibitem{HJ87}
P. G. Hansen and B. Jonson, Europhys. Lett. {\bf 4} (1987) 409.
\bibitem{Simon99}
H. Simon et al., Phys. Rev. Lett. {\bf 83} (1999) 496.
\bibitem{Keller94}
H. Keller  et al., Z. Phys. {\bf A348} (1994) 61.
\bibitem{Iwa00}
H. Iwasaki et al., Phys. Lett. {\bf B491} (2000) 8.
H. Iwasaki et al., Phys. Lett. {\bf B481} (2000) 7.
\bibitem{Mg32}
D. Guillemaud-M\"oller et al., Nucl. Phys. {\bf A426} (1984) 37.\\
T. Motobayashi et al.,  Phys. Lett. {\bf B346} (1995) 9.
\bibitem{Ozawa00}
A. Ozawa et al., Phys. Rev. Lett. {\bf 84} (2000) 5493.
\bibitem{HF}
 X. Campi, H. Flocard, A. K. Kerman and S. Koonin, Nucl. Phys. {\bf A251}
   (1975) 193;\\
 I. Hamamoto, H. Sagawa and X. Z. Zhang, Phys. Rev. 
{\bf C53} (1996) 765; \\
J. Dobaczewski et al.,   Phys. Rev. {\bf C53} (1996) 2809;\\
J. Meng and P. Ring, Phys. Rev. Lett. {\bf 80} (1998) 460.
\bibitem{shell}
A. Poves and J. Retamosa, Phys. Lett. {\bf B184} (1987) 311;
 Nucl. Phys. {\bf A571} (1994) 221; \\
E. K. Warburton, A. Becker and  and B. A. Brown, Phys. Rev. {\bf C41} (1990)
  1147;\\
 S. E. Koonin, D. J. Dean and K. Langanke, Annu. Rev. Nucl. Part. Sci.
   {\bf 47} (1997) 463; \\
 Y. Utsuno, T. Otsuka, T. Mizusaki and M. Honma, Phys. Rev. {\bf C60} (1999)
054315.
\bibitem{BM75}
A. Bohr and B.R. Mottelson, {\it Nuclear Structure}, vol. II (Benjamin,
New York, 1975).
\bibitem{Mah85}
C. Mahaux, P.F. Bortignon, C.H. Dasso and R.A. Broglia, Phys. Rep. {\bf 120} 
(1985) 1, and references therein. 
\bibitem{Colo01}
 G. Colo, Toshio Suzuki and  H. Sagawa, Nucl. Phys.  {\bf A695}, 167 (2001).
\bibitem{HSZ97}
 I.Hamamoto, H.Sagawa and X.Z.Zhang, Nucl. Phys. {\bf A626}, 669 (1997).
\bibitem{SA01}
 H.Sagawa and K. Asai,  Phys. Rev. {\bf C63}, 064310 (2001).
\bibitem{note}
When calculating the matrix elements of $H_{PV}$, the momentum dependent
terms are dropped  from  $v(r)$. 
\bibitem{VG80}
V. Bernard and Nguyen Van Giai, Nucl. Phys. {\bf A348} (1980) 75.
\bibitem{NVM}
N. Vinh Mau, Nucl. Phys. {\bf A592} (1995) 33. 
\bibitem{Nun96}
F.M. Nunes, I.J. Thompson and R.C. Johnson, Nucl. Phys. {\bf A596} (1996) 171.
\bibitem{BM}
A. Bohr and B.R. Mottelson, {\it Nuclear Structure}, vol. I (Benjamin,
New York, 1969). 
\bibitem{AS}
F. Ajzenberg-Selove, Nucl. Phys. {\bf A490} (1988) 1.
\bibitem{raman87}
 S. Raman et al., At. Data Nucl. Data Tables  36(1987) 1.

\end{thebibliography}
\end{document}